\documentclass[12pt]{elsarticle}

\usepackage{amssymb}
\usepackage{caption}
\usepackage{deluxetable}
\usepackage{multirow}
\usepackage{todonotes}
\journal{NIM A}

\begin{document}

\begin{frontmatter}



\title{Characterizing and correcting electron and hole trapping in germanium cross-strip detectors}


\author[a]{Sean N. Pike}
\author[a,b]{Steven E. Boggs}
\author[b]{Jacqueline Beechert}
\author[a]{Jarred Roberts}
\author[c]{Albert Y. Shih}
\author[b]{John A. Tomsick}
\author[b]{Andreas Zoglauer}

%\affiliation[inst1]{organization={Center for Astrophysics \& Space Sciences, University of California, San Diego},%Department and Organization
%            addressline={9500 Gilman Drive}, 
%            city={La Jolla},
%            postcode={92093}, 
%            state={CA},
%            country={USA}}

\affiliation[a]{organization={Department of Astronomy \& Astrophysics, University of California, San Diego},%Department and Organization
            addressline={9500 Gilman Drive}, 
            city={La Jolla},
            state={CA},
            postcode={92093}, 
            country={USA}}
            
\affiliation[b]{organization={Space Sciences Laboratory, University of California, Berkeley},%Department and Organization
            addressline={7 Gauss Way}, 
            city={Berkeley},
            state={CA},
            postcode={94720}, 
            country={USA}}
\affiliation[c]{organization={NASA Goddard Space Flight Center},
city={Greenbelt}, state={MD}, postcode={20771}, country={USA}}

\begin{abstract}

We present measurements of electron and hole trapping in three COSI germanium cross-strip detectors. By characterizing the relative charge collection efficiency (CCE) as a function of interaction depth, we show that intrinsic trapping of both electrons and holes have significant effects on the spectroscopic performance of the detectors. We find that both the electron and hole trapping vary from detector to detector, demonstrating the need for empirical trapping measurements and corrections. Using our measurements of charge trapping, we develop a continuous depth-dependent second-order energy correction procedure. We show that applying this empirical trapping correction produces significant improvements to spectral resolution and to the accuracy of the energy reconstruction.

\end{abstract}


% \begin{highlights}
% \end{highlights}

\begin{keyword}
Germanium semiconductor detectors \sep charge trapping \sep gamma-ray spectroscopy
% %% MSC codes here, in the form: \MSC code \sep code
% %% or \MSC[2008] code \sep code (2000 is the default)
\end{keyword}

\end{frontmatter}

% \linenumbers

\section{Introduction}
\label{sec:intro}

Germanium detectors (GeD) instrumented with pixel- or strip-geometry electrodes and readout electronics detect individual photon interactions with excellent spectral, spatial, and temporal resolution in the keV to MeV range. This capability allows for complex scientific analyses of high-energy astrophysical sources such as accreting black holes, supernova remnants, and more. Furthermore, incoming photons may undergo Compton scattering in the bulk of the GeD crystal, preferentially scattering in the direction of the photon's polarization vector. If the photons scattered in this way can be detected in addition to the initial interaction, then information about source polarization may be inferred. The Compton Spectrometer and Imager (COSI) will take advantage of this technique by utilizing stacked GeDs in order to measure multiple interaction sites for each incoming photon \cite{COSI}, allowing the observatory to gather spectral, timing, and polarization data for a plethora of sources in the 0.2-5\,MeV sky. COSI is a NASA Small Explorer (SMEX) mission scheduled to be launched in 2027 which will build on technologies developed for The Nuclear Compton Telescope (NCT) and the initial COSI payload developed as part of NASA's APRA program. These balloon experiments underwent an extensive flight program between 2005 and 2020 \cite{Boggs2004,Chiu2015,Kierans2016}. Throughout this paper, we will refer to the COSI balloon mission as COSI-APRA and to the upcoming space mission as COSI-SMEX.

GeDs operate via the Shockley-Ramo theorem whereby the energy deposited into the germanium crystal by each photon interaction is inferred from the current induced on the electrodes by the motion of charge carriers (i.e. liberated electrons and their corresponding positively-charged holes) across an electric potential. The induced charge on each electrode may be parameterized through the so-called ``weighting potential" \cite{Shockley1938,Ramo1939}. The contribution of each drifting charge carrier to the induced signal is proportional to the difference in weighting potential at its start and end positions. Therefore, if charge is prevented from moving through the detector, the induced signal will be suppressed. This can occur due to impurities, dislocation, or disordered regions, which attract and trap charge carriers. Impurities and dislocations are introduced in the process of crystal growth \cite{Hansen1971,Glasow1976}. Impurities may preferentially contribute to the trapping of electrons in p-type detectors \cite{Mei2020}. Disordered regions, on the other hand, are produced when the crystal is damaged by high-energy neutrons and protons. These regions develop a negative charge, thereby forming hole traps \cite{Goulding1972,Pehl1977,Darken1980}. Because the number of charge traps encountered by a cloud of charge carriers depends on the distance of the initial photon interaction from the corresponding collecting electrode, the overall effect of charge trapping is to degrade the energy resolution of the detector when spectra are produced using events at all depths. Measuring and correcting the effect of trapping on charge collection efficiency (CCE) is therefore an important step in optimizing the spectroscopic performance of GeDs (see, e.g., \cite{Hull2014,Arnquist2023}). This is especially true for missions designed to operate in space where energetic particles, such as $>20$\,MeV protons \cite{Ginet2013}, continually damage the detectors, leading to increased charge trapping over time.

In this paper, we present preliminary measurements of intrinsic charge trapping in GeDs which flew on board the COSI-APRA balloon missions, building upon the work presented in \cite{Beechert2023}. While previous studies of charge trapping in GeDs have measured the combined effects of electron- and hole-trapping \cite{Hull2014}, we are able to disentangle these effects and measure each of them separately with the novel readout scheme of the COSI-APRA detectors. In Section \ref{sec:data}, we briefly describe the detectors as well as the data and first-order calibration pipeline which produced the spectra we use for measurements of CCE. In Section \ref{sec:trapping}, we present the methods and results of our charge trapping study, and in Section \ref{sec:corrections} we present the improvements to spectral resolution which can be achieved using a simple empirical trapping correction technique. Finally, we summarize our results and conclusions in Section \ref{sec:summary}.

\section{COSI Detectors and Calibration}
\label{sec:data}

\begin{figure}
    \centering
    \includegraphics[scale=0.3]{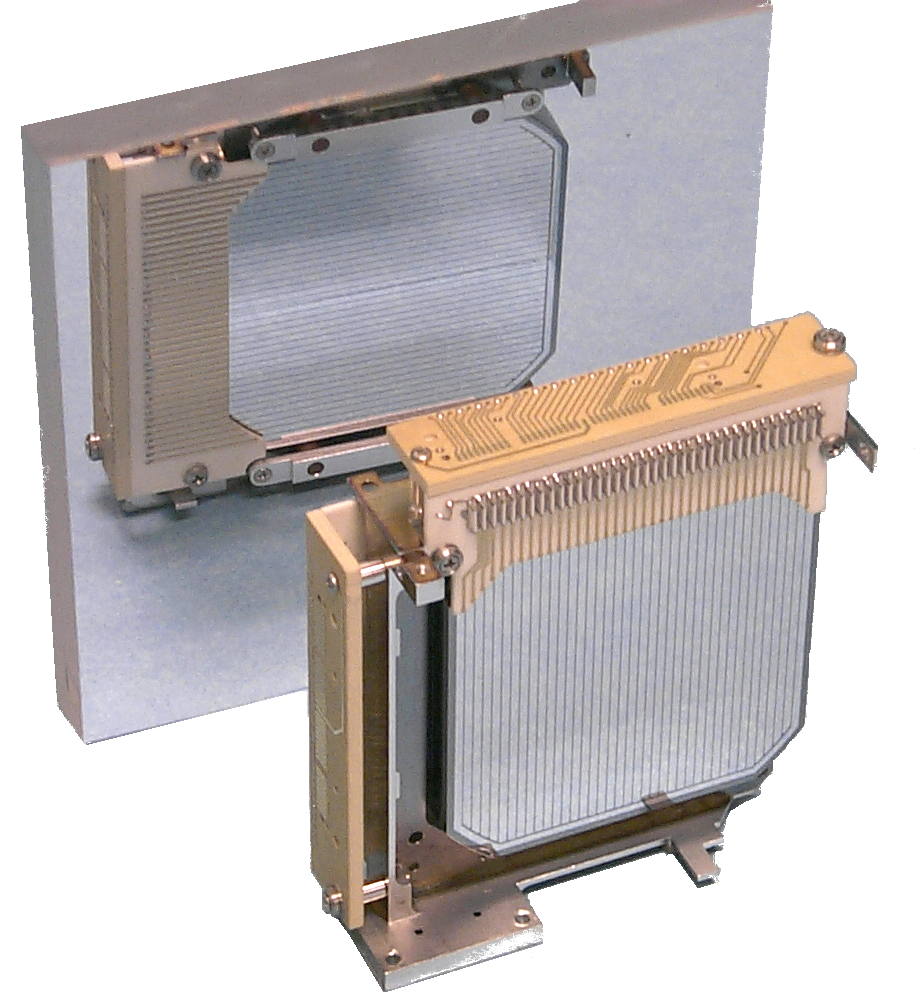}
    \caption{: A sample COSI-APRA GeD. Each detector, consisting of high-purity germanium, is instrumented with 37 orthogonal strip-geometry electrodes on the anode and cathode sides. The orthogonal strips on the opposing face of the GeD are shown using a mirror.}
    \label{fig:det_pic}
\end{figure}

\begin{figure}
    \centering
    \includegraphics[scale=0.69]{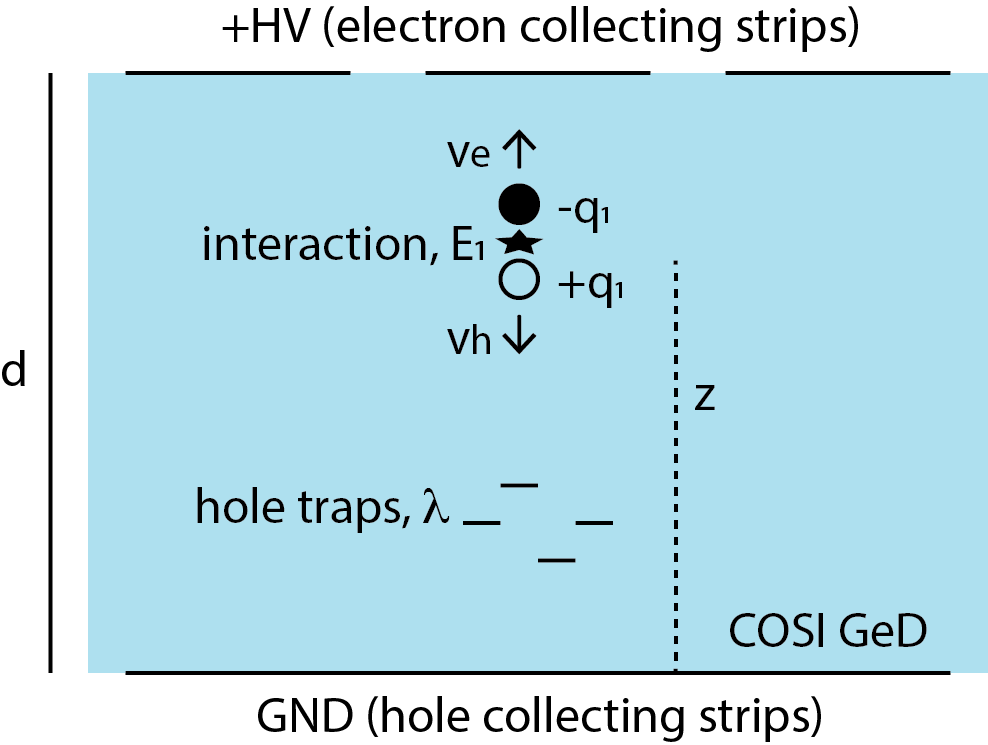}
    \includegraphics[scale=0.69]{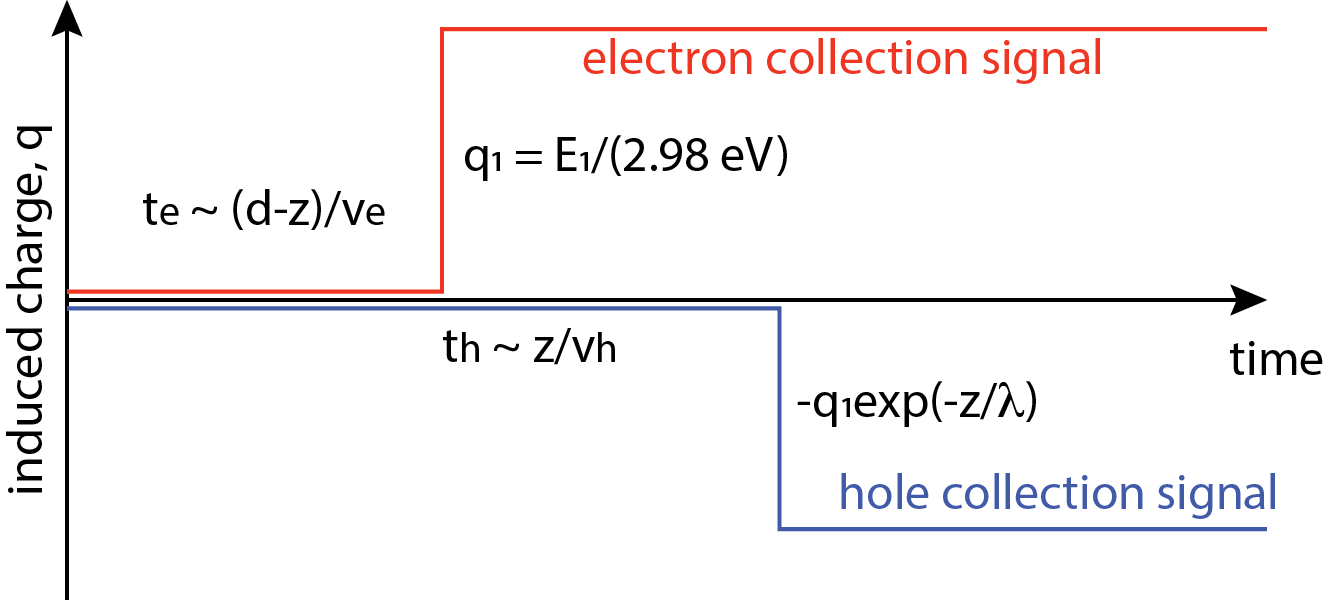}
    \caption{: Diagrams illustrating the relation between interaction depth and charge carrier drift length. Left: a diagram of the detector bulk. Right: a sketch of the charge induced on each side of the detector as a function of time. The further from a given collecting electrode that a photon interacts with the crystal, the longer it takes for the corresponding charge carriers to reach that electrode. The longer the electron drift length, the shorter the hole drift length, and vice versa. Therefore, the depth of interaction may be inferred by measuring the difference in collection time between the two types of charge carriers.}
    \label{fig:depth_diagram}
\end{figure}

The COSI-APRA instrument consisted of 12 high-purity GeDs. These detectors are custom, large-volume (54\,cm$^2$ area, 1.5\,cm thick) cross-strip germanium detectors utilizing amorphous contact technologies \cite{amman2007amorphous}. A sample detector is shown in Figure \ref{fig:det_pic}. Orthogonal electrodes on the opposite faces localize interaction sites in the x- and y-dimensions, and signal timing differences between the orthogonal strips provide localization in the z-direction (see Figure \ref{fig:depth_diagram}) with a resolution of 0.2\,mm \cite{Lowell2016}, resulting in full 3-dimensional position resolution for interactions within the detector.\footnote{Similar 3D event localization methods have been developed for Cadmium Zinc Telluride detectors as well \cite{Abbene2020,Abbene2022}} In this work we are focused on our original 2.0-mm strip pitch GeDs (37 strip electrodes on each face) that flew on the COSI-APRA balloon payload.

The COSI-APRA detectors underwent calibration and benchmarking campaigns before launch in 2014 and 2016. The data we present in this paper originated from the 2016 calibration campaign. These calibrations, which are described in detail in \cite{Beechert2022}, are critical to the accurate reconstruction of 3-dimensional interaction position, time of arrival, and photon energy. For the purposes of the investigation we present in this paper, measurements of depth-of-interaction and of energy are the most relevant.

During the 2016 calibration campaign, the detectors were irradiated with seven different radioactive sources, yielding 15 different spectral lines covering an energy range of 60\,keV to 1836\,keV. Charge transport simulations were performed in order to produce templates of the distribution of collection time difference (CTD) between the electron and hole signals measured on opposite sides of the detector. For each pixel, an empirical CTD was measured using a $^{137}$Cs source, and the simulated data was benchmarked against the measured data in order to calibrate the relation between the difference in measured collection time and the interaction depth. 

In order to calibrate photon energy reconstruction, for each strip, the pulse heights produced by the 15 spectral lines were fitted to a third-order polynomial which related these measurements to the known energies of the lines. Each of the photopeaks were fitted to a Gaussian plus a linear background model to determine its centroid. The photopeaks were not fitted as a function of depth at this stage of calibration, therefore the effects of charge trapping on the CCE remain in the calibrated data. Additional energy calibrations, including corrections for temperature-dependent effects and for cross-talk with neighboring strips, were also applied. In this work we will refer to the 2016 energy calibration as the first-order energy calibration, while our work introduces second-order (depth-dependent) corrections.

\section{Measurement of Charge Trapping}
\label{sec:trapping}

In order to measure the effects of charge trapping on CCE, we investigated calibration data from the COSI-APRA detectors. Specifically, we utilized a data set which was produced during the 2016 calibration campaign by irradiating the instrument using a $^{137}$Cs source, which produces emission at 661.657\,keV via radioactive decay. For each detector, we applied the first-order calibration pipeline described in Section \ref{sec:data} in order to produce a list of all single-pixel events. Each event is tagged with the inferred 3-dimensional position of the interaction site as well as two values of the energy deposited as measured using the electron and hole signals. We included events from all strips, and we excluded events which triggered on multiple pixels.

Next, we binned each of the calibrated and filtered event lists according to depth of interaction. Each detector has a thickness of approximately $1.5$\,cm, and we sorted events into 15 depth bins, resulting in bins with width of 1\,mm (several times larger than the spatial resolution in the z-direction). We subsequently analyzed the distributions of each of the two inferred energies, essentially splitting the event lists into two: one with energies inferred from the electron signal, and another with energies inferred from the hole signal. For each of these sets of data, we ignored events with energies, $E$, outside of the range $640\,\mathrm{keV} < E < 672\,\mathrm{keV}$. This range was chosen in order to adequately capture the line emission at 662\,keV as well as the low-energy tailing which results from sub-threshold charge sharing across adjacent strips.

For each depth bin, we fitted the data to the line model described in \cite{BoggsAndPike2023}. While the first-order calibration of the data fitted the calibration lines to a simple Gaussian distribution, charge sharing across adjacent strips produces significant low-energy tailing in the spectra. Our goal was to determine the centroid energy for the line emission at different depths, meaning that accurate line profile modeling is crucial to avoid biasing the centroid measurements. The model we used combines a Gaussian component, representing the fundamental response of the detector, with a power-law tail and a much shallower linear tail accounting for charge sharing due to diffusion and repulsion. The resulting model for number of photon counts (i.e. events) as a function of reconstructed energy, $E$, can be written

\begin{equation}\label{eq:line_profile}
    f(E) = Ae^\frac{-(E-E_0)^2}{2\sigma^2} + \left[Be^{\Gamma (E-E_0)} + C(1+m(E-E_0))\right]\times \left[1-\mathrm{erf}\left({\frac{E-E_0}{\sqrt{2}\sigma_t}}\right) \right]
\end{equation}

\noindent where $A$, $B$, and $C$ are the normalizations of the Gaussian, power-law, and linear components, respectively, $E_0$ is the centroid of the Gaussian component and $\sigma$ is its standard deviation, $\Gamma$ is the photon index of the power-law tail, and $m$ is the slope of the linear tail. The term in the second set of brackets represents the high-energy turnover of the tail components. The width of this turnover is allowed to differ from that of the Gaussian component and is given by $\sigma_t$. Figure \ref{fig:corrected_spectra} clearly illustrates the low-energy tailing resulting from charge sharing as well as several fits to Equation \ref{eq:line_profile}.

Simulations of charge sharing in \cite{BoggsAndPike2023} demonstrated that a number of model parameters --- namely, $\Gamma$, $C/B$, $m$, and $\sigma_t/\sigma$ --- do not vary significantly with interaction depth. While they may differ somewhat from detector to detector, or even on a sub-detector level, we found that freezing these parameters at the values determined by \cite{BoggsAndPike2023} resulted in good fits allowing us to minimize bias in our measurements of the centroid and width of the Gaussian component. In the future, we hope to obtain sufficient data to directly fit for these parameters and investigate their variability. 

With these parameters fixed, we defined a spectral probability density function, $P(E)$, by normalizing Equation \ref{eq:line_profile} to unity over the energy range $640\,\mathrm{keV} < E < 672\,\mathrm{keV}$, thereby eliminating the need to fit the overall normalization factor $A$, which depends only on the number of events collected and not on any relevant detector characteristics. We next used the Python package iminuit \cite{iminuit} to construct the corresponding cost function, defined as the negative log-likelihood given $N$ unbinned events. The likelihood function in this case is written

\begin{equation}\label{eq:likelihood}
    \mathcal{L} = \prod_i^N P(E_i)
\end{equation}

\noindent such that the cost function is given by the negative of the sum over all measured photon energies, $E_i$, of $\log{P(E_i)}$.

For each depth bin, we fitted the line profile model to the event lists for the electron and hole signals separately. We performed the fits by minimizing the cost function described above using iminuit, which utilizes the minimization algorithm described in \cite{James1975}. By fitting the unbinned data in this way while also freezing parameters according to \cite{BoggsAndPike2023}, we were able to limit the fitted parameters to $B$, $E_0$, and $\sigma$. In other words, we fitted for the Gaussian centroid and width as well as the overall extent of charge sharing as captured by the normalization of the tail components. We therefore determined the relative shifts in the line profiles with depth, representing the effects of electron and hole trapping on CCE, unbiased by the low-energy tails which arise due to charge sharing.

\begin{figure}
    \centering
    \includegraphics[scale=0.4]{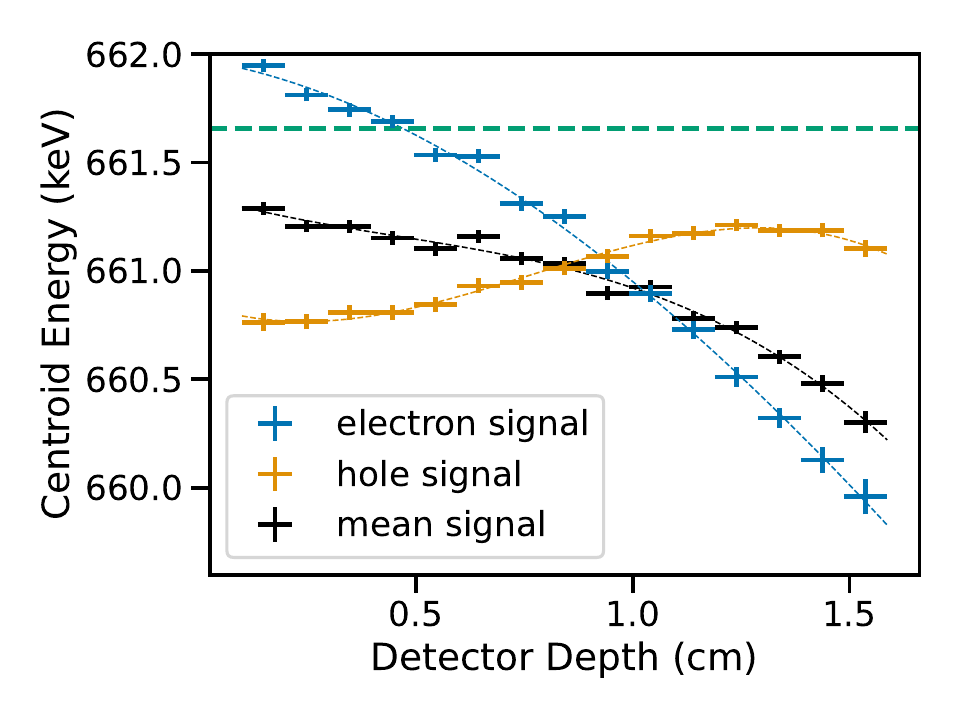}
    \includegraphics[scale=0.4]{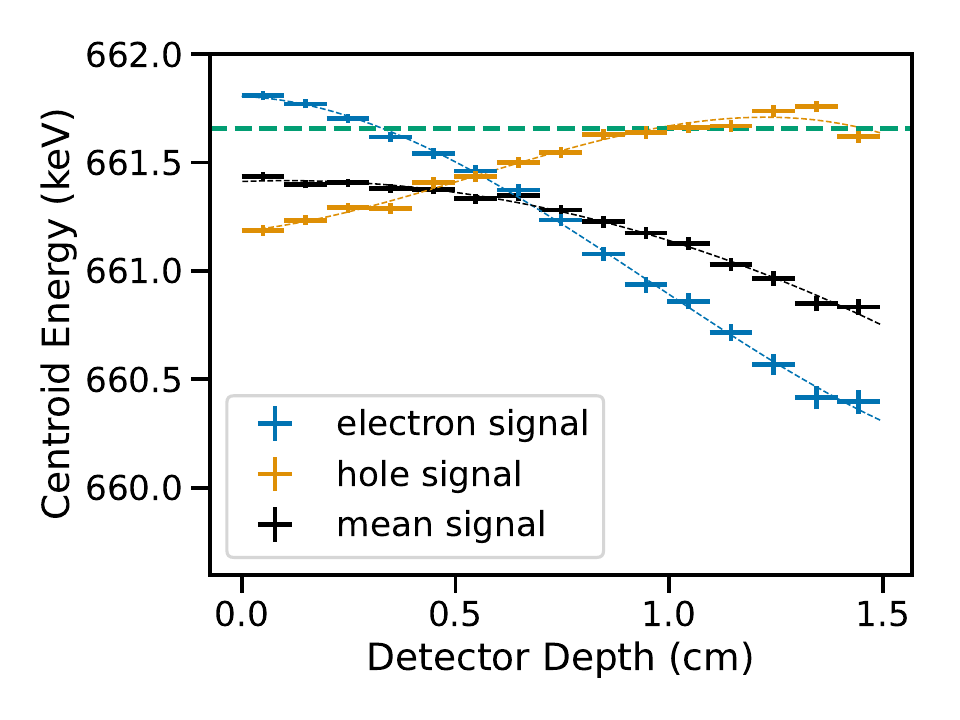}
    \includegraphics[scale=0.4]{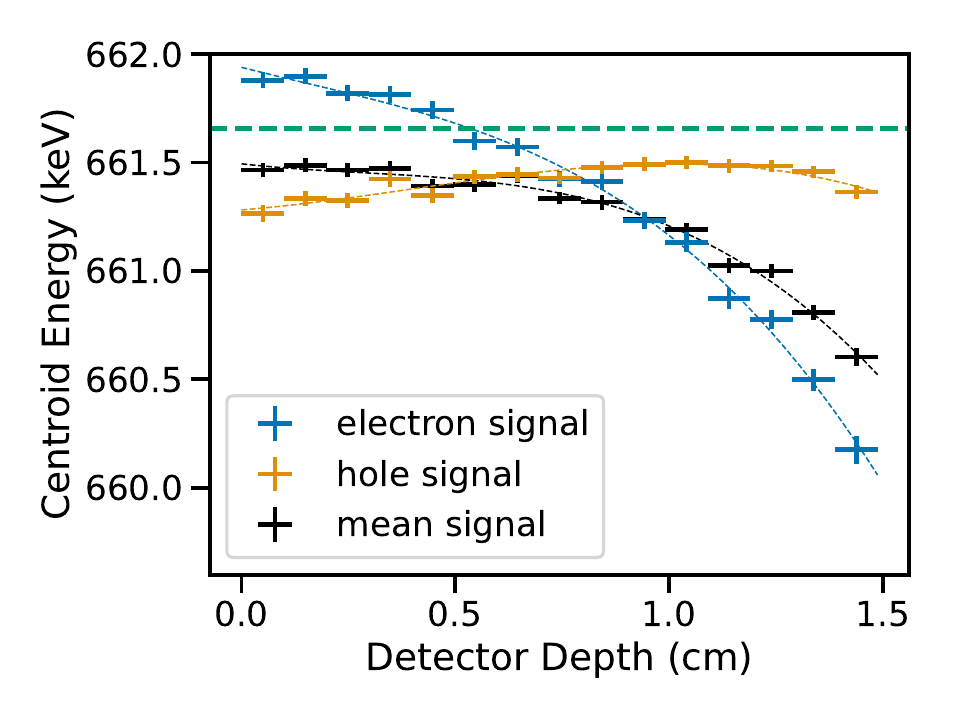}
    \caption{: Depth plots produced for a sample of three COSI-APRA detectors, Det2, Det4, and Det8, in clockwise order beginning from the top left. The centroid energies measured using only the electron signal are shown in blue, those measured using only the hole signal are shown in orange, and the centroid energies measured using event lists whose energies correspond to the mean of the electron and hole signals are shown in black. For each signal we produced a cubic spline, shown as a dashed line of the same color. The dashed green line shows the known energy of the $^{137}$Cs line at 661.657\,keV. As expected, the electron signal shows considerable depth dependence, indicating significant trapping. To a lesser extent, the hole signal also shows signs of measurable charge trapping, despite the fact that the detectors have not undergone significant radiation by energetic particles. The trapping evident in the mean signal is dominated by electron trapping. Analysis of the mean signal alone is not sufficient to determine the extent of hole trapping.}
    \label{fig:depth_plots}
\end{figure}

Here we present the results of this fitting procedure for three detectors, corresponding to Det2, Det4, and Det8 described in \cite{Kierans2018}. We chose to present these detectors in particular because their calibrated spectra were most accurately modeled using Equation \ref{eq:line_profile} and were not affected by issues like high-energy or low-energy excesses. The origin of the low-energy excess observed for some detectors, which manifests as a low-amplitude Gaussian component, is not yet known. The high-energy excess is likely due to insufficient calibration of cross-talk between adjacent strips, which amplifies some events. It is unlikely that either of these excesses are related to charge sharing, leading us to exclude these spectra. For each of the three chosen detectors, we plotted the measured value of $E_0$ using the first-order energy correction at each depth bin, where we defined depth as the distance from the electron-collecting side of the detector, such that increasing depth corresponds to increasing electron drift length and decreasing hole drift length. These depth plots are shown in Figure \ref{fig:depth_plots}. The centroid energies, $E_0$, measured for the electron signal are shown in blue while those measured using the hole signal are shown in orange. 

For all three detectors, it is apparent that trapping impacts the CCE as a function of interaction depth, and that electron trapping dominates over hole trapping. Across each detector, the line profile produced from the electron signal experiences a relative shift of around $2$\,keV, while the hole signal demonstrates a shift of less than $1$\,keV. If we approximate the percent of charge trapped as the centroid shift divided by the known line energy, this corresponds to a trapping fraction of around $0.3\%$ for electrons and $0.1\%$ for holes. This percentage may be energy-dependent. Additional charge trapping studies using different radioactive sources will help to determine whether the effects of charge trapping saturate at a higher energies or whether a consistent percentage of charge is trapped in the COSI bandpass. Furthermore, we find that each of the detectors we investigated exhibits unique levels of electron and hole trapping, demonstrating the importance of measuring actual trapping effects on a detector-by-detector basis. 

We note that for Det2 and Det8, the centroid energy measured for the hole signal using the first-order energy calibration never crosses the true value of $661.657$\,keV. The precise reason for this is not yet confirmed, but it is likely due to the method of first-order energy calibration applied to the data prior to our fitting procedure. For the purpose of those calibrations, the line profiles were fitted to a simple Gaussian component, which may bias the inferred energy given the significance of charge sharing effects. In addition, the energy-dependent gain curve determined using sources emitting at various energies may underestimate the photon energy in certain energy ranges depending on the accuracy of the polynomial fit. These are aspects of the first-order detector calibration which we are currently investigating. We emphasize that for the purposes of this study, the relevant proxy for charge trapping is the relative change in recorded energy with depth rather than absolute energy measurements.

In Figure \ref{fig:depth_plots} we also show in black the effects of trapping measured by fitting line profiles to unbinned event lists whose energies correspond to the mean of the electron and hole signals. These curves are analogous to the charge trapping curves measured by \cite{Hull2014}. In plotting the three curves together, it is apparent that the extent of both electron and hole trapping may be easily underestimated when examining only the mean. In particular, the effects of hole trapping on the CCE may be overlooked entirely and the downward curves shown in black may erroneously be attributed entirely to electron trapping. In any case, the collection of curves shown in Figure \ref{fig:depth_plots} demonstrates that considerations of charge trapping and interaction depth should be taken into account during energy calibration. In the following section, we present one method of depth-dependent second-order energy calibration in order to illustrate the improvements that can be achieved.

\section{Applying the Depth Correction}
\label{sec:corrections}

In order to compensate for the effects of trapping on CCE, we performed an event-by-event depth-dependent second-order energy correction on top of the existing first-order calibrations. Having measured the line centroids for both the electron and hole signals across 15 bins spanning the $1.5$\,cm detector thickness, we then produced smooth energy-depth relationships based on these binned measurements. Using the Python package SciPy \cite{SciPy} we produced a cubic spline in order to interpolate and extrapolate each of the binned curves shown in Figure \ref{fig:depth_plots}. These splines are shown as dashed lines of the same color as their corresponding measurements. Next, for each event, we corrected both energy estimates (corresponding to electron and hole signals) using their respective splines, $S(z)$, such that the corrected energy can be written $E_\mathrm{i,corr}(z)= E_\mathrm{i,uncorr}(z) \frac{661.657\,\mathrm{keV}}{S(z)}$. We then fitted each of the lists of corrected electron-signal energies and hole-signal energies to the line profile model described by Equation \ref{eq:line_profile}. Additionally, we produced a third list of energies corresponding to the mean of these second-order-corrected electron- and hole-signal values. We also fitted this list of events to the line profile model. 

\begin{figure}
    \centering
    \includegraphics[scale=0.275]{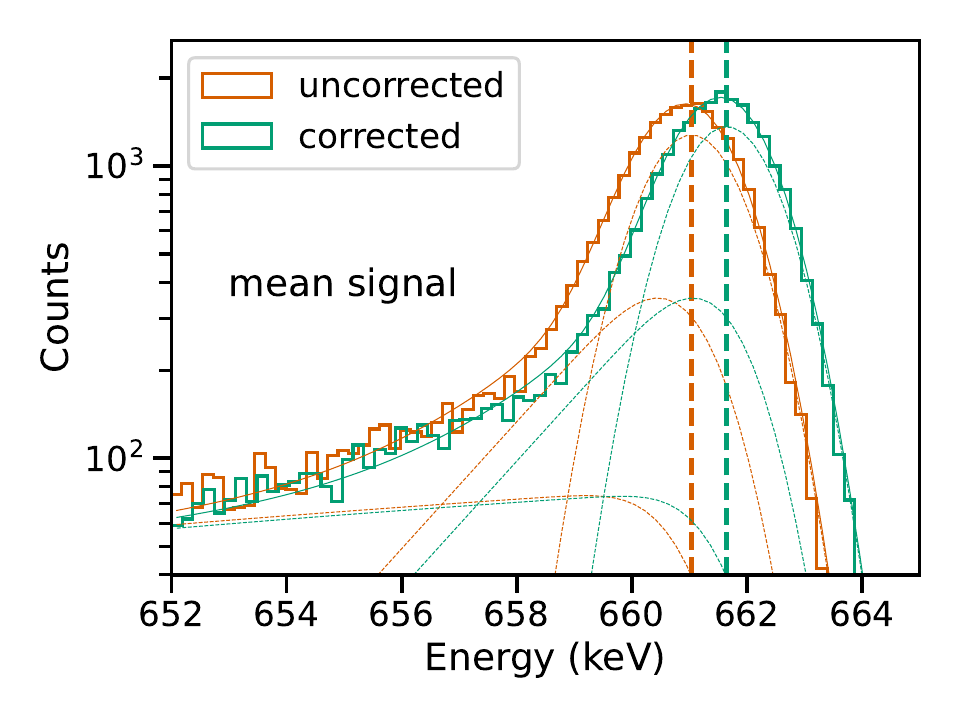}
    \includegraphics[scale=0.275]{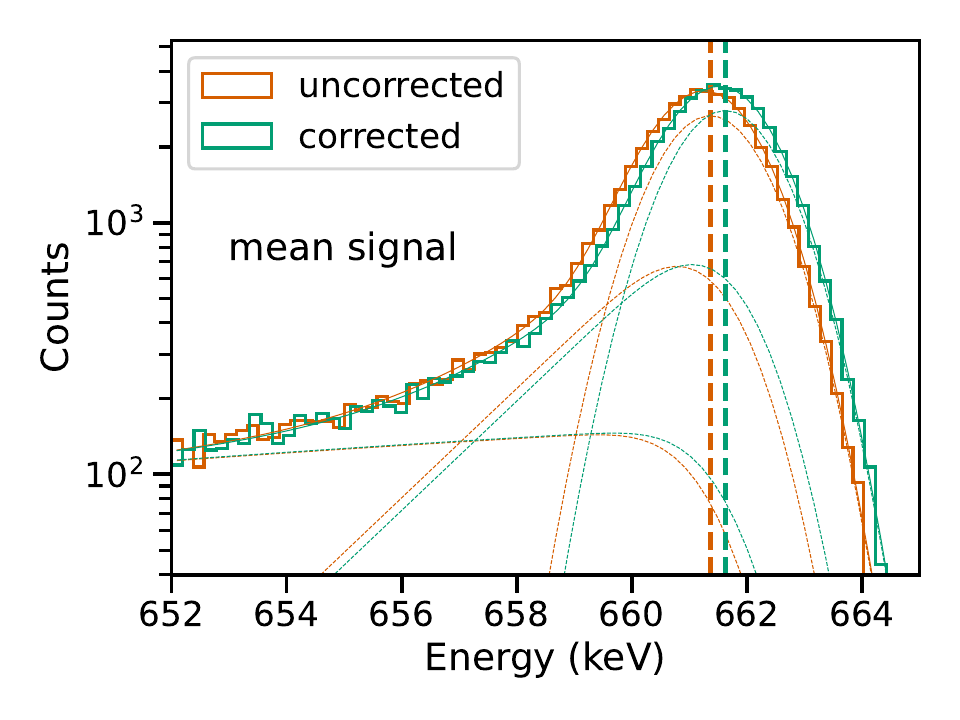}
    \includegraphics[scale=0.275]{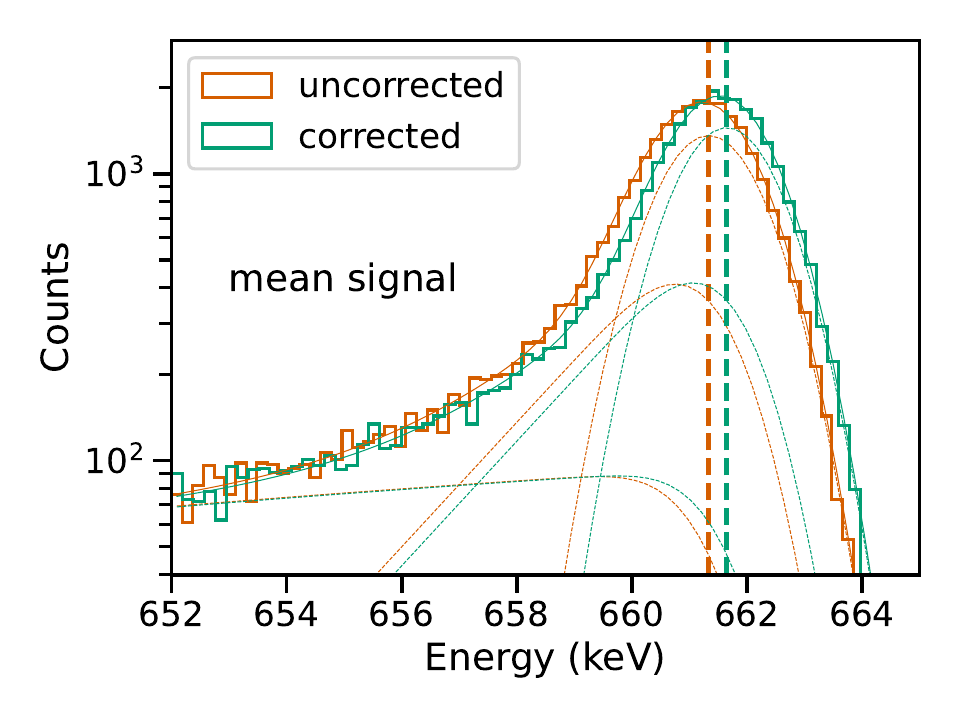}
    \includegraphics[scale=0.275]{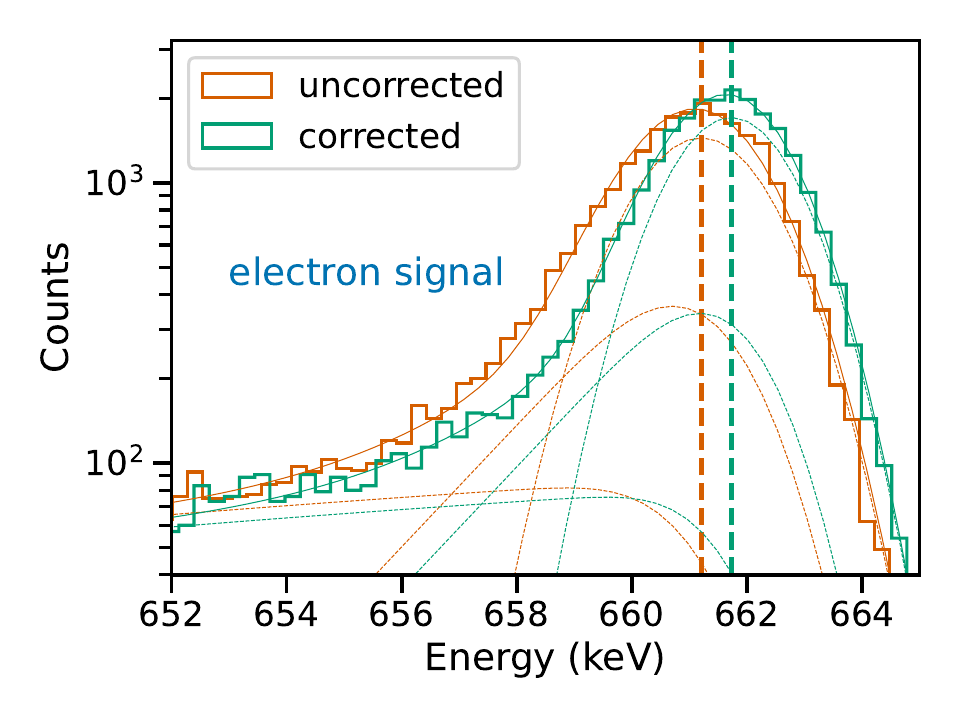}
    \includegraphics[scale=0.275]{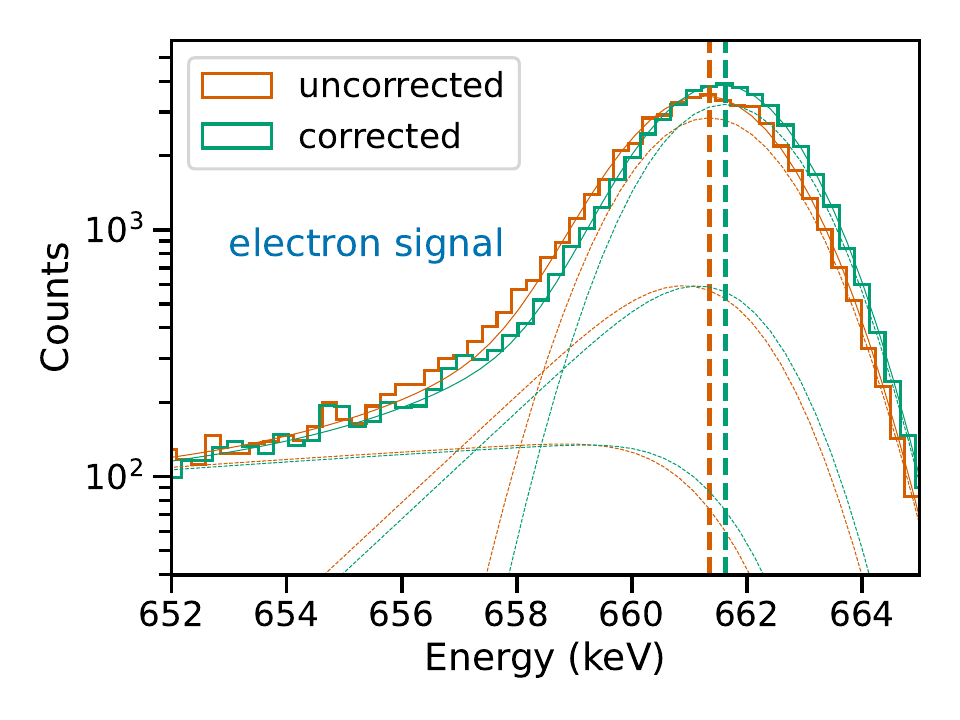}
    \includegraphics[scale=0.275]{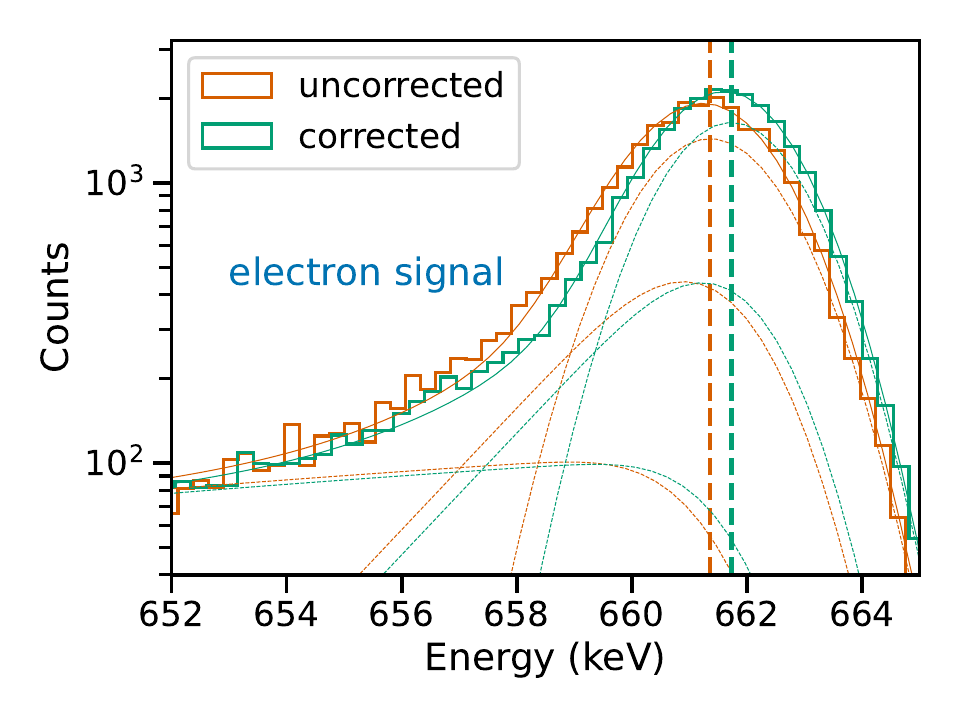}
    \includegraphics[scale=0.275]{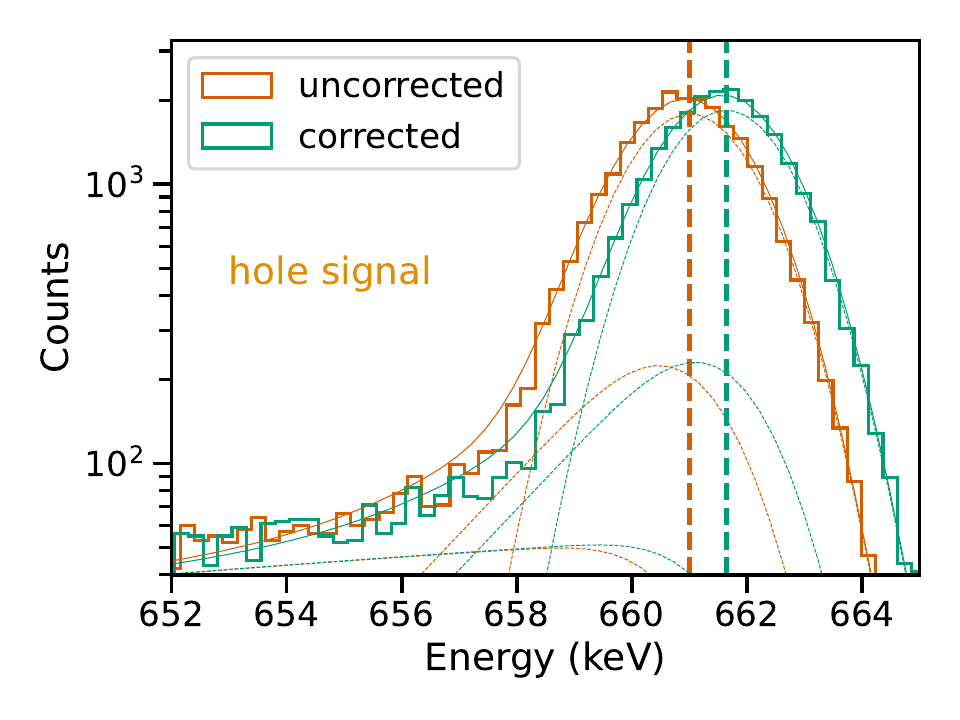}
    \includegraphics[scale=0.275]{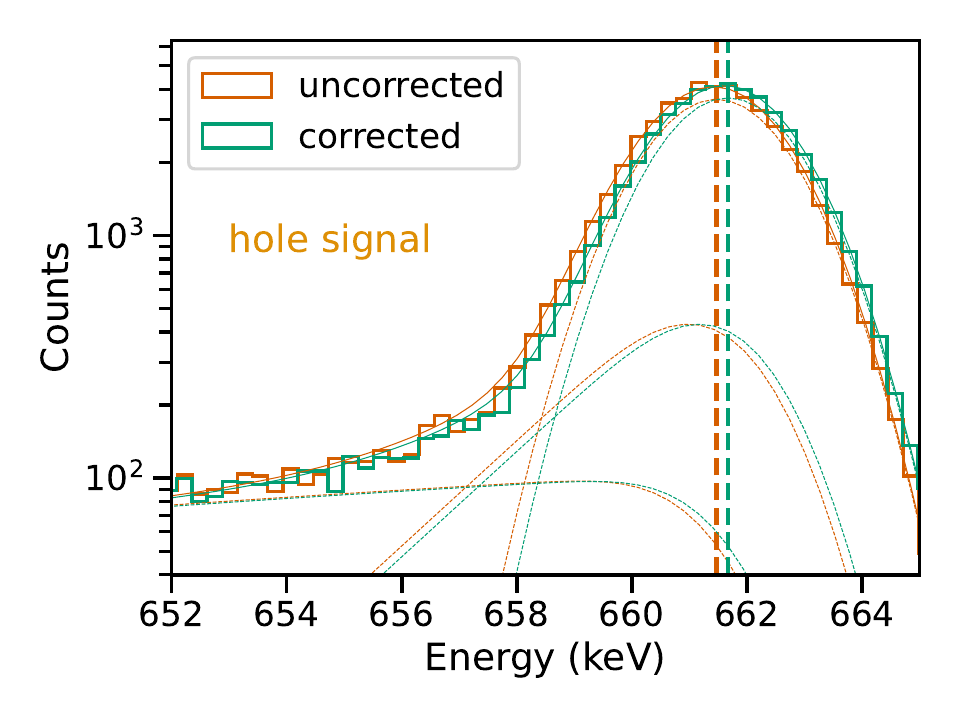}
    \includegraphics[scale=0.275]{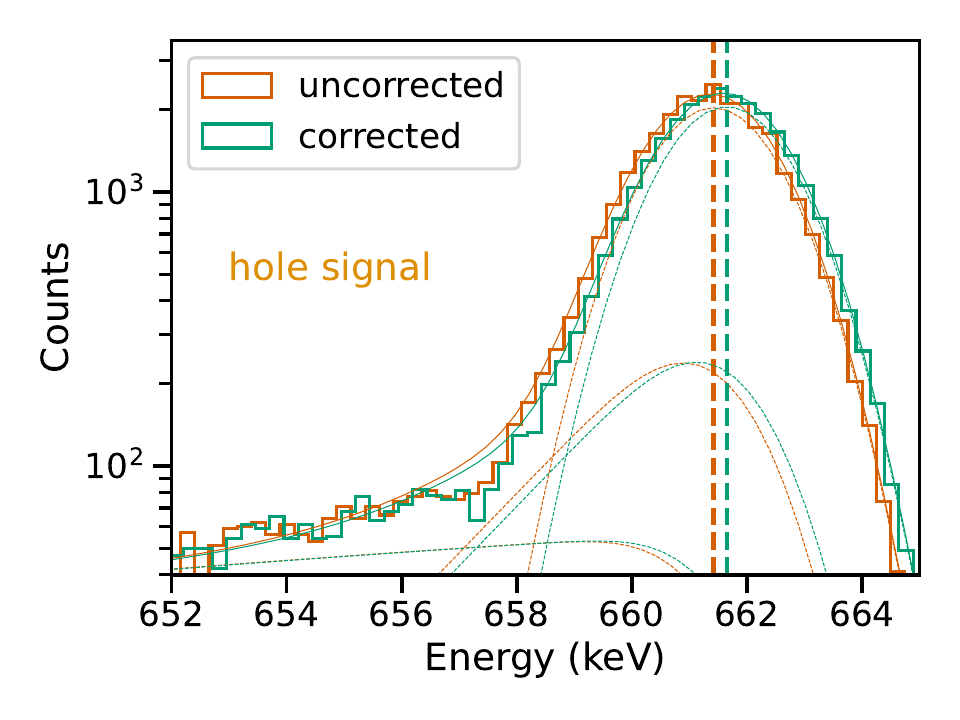}
    \caption{: Spectra produced before (orange) and after (green) application of depth correction for Det2 (left column), Det4 (middle column), and Det8 (right column). The spectra in the top row were constructed from events for which the energy was determined by averaging the energy inferred from the electron and hole signals, while the middle and bottom rows show the spectra constructed from only the electron and hole signals, respectively. Events at all depths in each detector were included. The finely dashed curves show the Gaussian, power-law tail, and linear tail components described in Equation \ref{eq:line_profile}, and the vertical dashed lines indicate the measured centroid energy for each spectra. Application of the depth correction results in a much more accurate energy determination and improvements in the spectral resolution by up to $\Delta\mathrm{FWHM}=0.43$\,keV for the mean signal.}
    \label{fig:corrected_spectra}
\end{figure}

We show the binned spectra, before and after depth corrections, in Figure \ref{fig:corrected_spectra}. In Table \ref{tab:corrections}, we list the spectral resolution, defined as the full-width at half-maximum (FWHM) of the Gaussian component, for the electron, hole, and mean spectra produced for the three detectors before and after depth corrections. We also list the change in the FWHM, defined in quadrature: $\Delta\mathrm{FWHM}=\sqrt{\mathrm{FWHM^2_{uncorr} - FWHM^2_{corr}}}$, and we list the logarithmic difference between the measured centroid and the actual line energy before and after corrections. We find that for the spectra produced by the electron signal, the resolution is improved significantly, with $\Delta\mathrm{FWHM}$ of around $1$\,keV. Because the extent of hole trapping is far less than that of electron trapping, resolution improvements produced by our depth correction are smaller but still significant for Det2 and Det4, with $\Delta\mathrm{FWHM}\approx400$\,eV. The picture is less clear for Det8, which exhibits a much weaker hole trapping signal and for which $\Delta\mathrm{FWHM}$ is not significant. Nonetheless, when comparing the spectra produced using the mean of the electron and hole signals, we find consistent improvements to spectral resolution across all three detectors following application of the second-order energy calibration. 

Notably, the application of depth corrections brings the measured line centroids into much closer agreement with the actual line energy. We calculated the logarithmic relative difference between the measured centroid and the true energy, $\log_{10}\left(1-\frac{E_0}{661.657\,\mathrm{keV}}\right)$, which we list in the rightmost two columns in Table \ref{tab:corrections}. The application of our second-order energy calibrations decreased the relative difference between the derived line energy and the true energy by a factor of up to 100 compared to the application of only the first-order calibrations.

Finally, we determined the trapping curve for each detector using only the mean signal --- i.e., the event energies inferred by taking the mean of the first-order calibrated electron and hole signals --- shown in black in Figure \ref{fig:depth_plots}. We again produced a cubic spline which we used to correct the mean signals on an event-by-event basis. In short, rather than applying our second-order energy calibration prior to calculating the mean of the electron and hole signals, we applied this correction to the mean signal calculated as the average of the uncorrected electron and hole signals. We found that for all three detectors, this method resulted in similar, but somewhat less effective improvements to the spectral resolution, indicating that applying the second-order correction to the electron and hole signals independently is the preferred method.

\begin{deluxetable}{ccccccc}
    \tablenum{1}
    \tablecaption{Spectral resolution and relative centroid shifts before and after application of second-order depth-dependent corrections for a sample of 3 COSI-APRA detectors. The row labeled ``mean" refers to the spectra constructed from the mean of the second-order corrected electron and hole signals. \label{tab:corrections}}
    \tablewidth{0pt}
    \tablehead{ \colhead{Detector} & \colhead{Signal} & \multicolumn{2}{c}{FWHM} & \colhead{$\Delta$FWHM} & \multicolumn{2}{c}{$\log_{10}\left(1-\frac{E_0}{661.657\,\mathrm{keV}}\right)$} \\
                    &        & \multicolumn{2}{c}{(keV)}  & \colhead{(eV)} &  &   \\
                    &        & \colhead{Uncorr.}   & \colhead{Corr.}   &  & \colhead{Uncorr.}  & \colhead{Corr.}}
    \startdata
    \multirow{3}{*}{Det2}   & electron  & $2.85(2)$  & $2.61(2)$    & $1160 \pm 60$  & $-3.17(1)$  & $-3.96(6)$\\
                            & hole      & $2.69(2)$  & $2.66(2)$    & $400 \pm 140$ & $-3.006(6)$  & $-4.6(2)$\\
                            & \textbf{mean} & $2.12(2)$ & $2.08(1)$     & $410 \pm 110$ & $-3.030(7)$  & $-4.9(4)$\\
                            \hline
    \multirow{3}{*}{Det4}   & electron  & $3.12(1)$  & $2.99(1)$  & $900 \pm 70$   & $-3.33(1)$   & $-4.3(1)$\\
                            & hole      & $2.92(1)$  & $2.88(1)$  & $450 \pm 100$  & $-3.56(2)$   & $-4.6(2)$\\
                            & \textbf{mean} & $2.28(1)$  & $2.26(1)$  & $240 \pm 150$   & $-3.35(1)$   & $-4.32(9)$\\
                            \hline
    \multirow{3}{*}{Det8}   & electron  & $3.05(2)$  & $2.87(2)$  & $1020 \pm 80$   & $-3.35(2)$   & $-3.97(7)$\\
                            & hole      & $2.72(1)$  & $2.72(1)$  & $120 \pm 450$  & $-3.45(2)$   & $-5.1(7)$\\
                            & \textbf{mean} & $2.23(2)$  & $2.18(2)$  & $430 \pm 110$  & $-3.31(1)$   & $-5.0(6)$ 
    \enddata
\end{deluxetable}

\section{Summary and Conclusions}
\label{sec:summary}

We have presented measurements of the effects of charge trapping on CCE in GeDs for both negative and positive charge carriers simultaneously, made possible by the readout design of the COSI-APRA detectors. Using line profile models developed in \cite{BoggsAndPike2023} which accurately account for low-energy tailing due to charge sharing, we were able to accurately and precisely measure the apparent shift in spectral lines due to incomplete charge collection at larger drift distances, corresponding to different depths in the detectors. Having measured the effect of charge trapping in this way, we were also able to apply an empirical depth-dependent second-order energy  correction to each event, bringing the inferred energies into better agreement with the actual energies of interacting photons.

Our measurements showed that although the detectors under investigation have been exposed to far less high-energy particle flux than that expected during the space mission, significant hole trapping is still evidenced by variations in the CCE with depth. Correcting for depth-dependent hole trapping provided significant improvements in spectral resolution for two of the three detectors under investigation. As expected, we found that electron trapping is a much stronger effect in these detectors. Furthermore, we showed that the electrons are trapped at a much higher rate than is inferred when only using the mean of the electron and hole signals. This finding demonstrates the importance of measuring trapping from both sides of the detector and contradicts the assumption that undamaged detectors only exhibit intrinsic electron trapping.

Charge trapping degrades the spectral resolution of GeDs when events at all depths are used to reconstruct spectra. Beginning from our measurements of charge trapping, we produced an empirical depth correction procedure by which the energy inferred from the electron and hole signals for each event using the first-order energy calibrations were corrected by a continuous depth-dependent factor. By bringing the centroids of the measured spectral lines into much better agreement with the actual energy, these corrections demonstrated the importance of utilizing line profile models which more accurately match the shape of the observed profile, as well as the utility of applying depth corrections in addition to the existing suite of first-order energy calibrations. Applying this correction led to significant improvements in the spectral resolution of $\Delta\mathrm{FWHM/FWHM}\geq11\%$. For spectra measured using only the electron signals, the improvements were much more noticeable, $\Delta\mathrm{FWHM/FWHM}\geq29\%$. This result suggests that as the effects of hole trapping increase due to damage over the lifetime of these and similar detectors, the overall spectral resolution improvements that can be gained from this kind of depth correction procedure will become increasingly critical to preserving optimal spectral performance. Additionally, the empirical nature of this correction method makes it applicable to any detector for which the photopeak centroid for each charge carrier can be reliably measured as a function of depth.

We presented these results for three COSI-APRA detectors, and we found that the severity of both hole and electron trapping differed significantly from detector to detector, illustrating the importance of characterizing trapping in individual detectors. Furthermore, this result suggests that trapping may even differ on a sub-detector level if there are significant gradients in the density of charge traps across individual detectors. By measuring the charge trapping for individual strips or even individual pixels, we may probe impurities and other sources of charge traps as a function of x- and y- position, thereby complementing other methods of impurity measurement (e.g. \cite{Abt2023}). This is an area of investigation which we reserve for future work.

Unlike COSI-APRA, COSI-SMEX will fly in low-Earth orbit, where damage to the detectors due to high-energy protons will significantly increase the rate of hole trapping in the GeDs. The work we have presented here will help to lay the groundwork for understanding and correcting trapping effects throughout the COSI-SMEX mission. Further investigations, such as developing predictions for the evolution of charge carrier trapping throughout the lifetime of the mission, modeling the effects of trapping on the shape of measured line profiles, and using simulations to link physical crystal properties to measurements of charge trapping, will be necessary to fully account for the effects of charge trapping in the COSI-SMEX GeDs.

\section{Acknowledgements}
\label{}

This work was supported by the NASA Astrophysics Research and Analysis (APRA) program, grant 80NSSC21K1815. We thank the anonymous referees whose comments and suggestions improved the quality of this work.

 \bibliographystyle{elsarticle-num} 
 \bibliography{main}





% \end{thebibliography}
\end{document}